\documentclass[twocolumn,10pt]{asme2e}

\usepackage{amssymb,amsmath}
\usepackage{graphics} 
\usepackage{graphicx}
\usepackage{subfigure}
\usepackage{enumerate}
\usepackage{multirow, multicol}
\usepackage[english]{babel}
\usepackage{cite}
\usepackage{nomencl}
\makenomenclature
\makeindex
\usepackage{cleveref}
\usepackage[usenames,dvipsnames]{color}
\usepackage{csquotes}
\usepackage{mathrsfs}
\usepackage{upgreek}
\usepackage{mathtools, cuted}

\usepackage{algorithm}
\usepackage{algpseudocode}

\usepackage{epstopdf}
\usepackage{float}
\usepackage{subfigure}

\usepackage{tikz}
\usetikzlibrary{shapes,arrows,calc}

\usepackage{enumerate,enumitem}
\usepackage{multirow, multicol}
\usepackage[english]{babel}

\usepackage[usenames,dvipsnames]{pstricks}
\usepackage{epsfig}
\usepackage{pst-grad} 
\usepackage{pst-plot} 

\newtheorem{theorem}{Theorem}

\usepackage{algorithm}
\usepackage{algpseudocode}

\confshortname{DSCC2019}
\conffullname{the ASME 2019 \\ Dynamic Systems and Control Conference}

\confdate{October 8-11}
\confyear{2019}
\confcity{Park City, UT}
\confcountry{USA}

\papernum{DSCC2019-9098}

\title{Delay-Dependent Output-Feedback Control for Blood Pressure Regulation Using LPV Techniques}

\author{Shahin Tasoujian\thanks{Shahin Tasoujian is a Ph.D. candidate in Mechanical Engineering at the University
of Houston, Houston, TX 77204, U.S.A.  e-mail: stasoujian@uh.edu.
\: \: \: \: K. Grigoriadis and M. Franchek are also with the University of
Houston.},  Karolos Grigoriadis, and Matthew Franchek
   
    \affiliation{
	Department of Mechanical Engineering\\
	University of Houston\\
	Houston, Texas 77204\\
    Email: stasoujian@uh.edu
    }	
}

\begin{document}

\tikzset{
block/.style = {draw, fill=white!20, rectangle, thin, minimum height=2.0em, minimum width=1.50em, text centered},
tmp/.style  = {coordinate}, 
sum/.style= {draw, fill=white!20, circle, node distance=.5cm},
tip/.style = {->, >=stealth', thin, dashed},
input/.style = {coordinate},
output/.style= {coordinate},
pinstyle/.style = {pin edge={to-,thin,black}}
}

\maketitle    
\begin{abstract}
{\it This paper presents a delay-dependent parameter-varying control design approach to address the automated blood pressure regulation problem in the critical patient resuscitation using closed-loop administration of vasopressors. The mean arterial pressure (MAP) response of a patient subject to the intravenous vasoactive drug treatment is modeled as a linear parameter-varying (LPV) model, where varying model parameters and varying time-delay are considered as scheduling parameters of the system. Parameter-dependent Lyapunov-Krasovskii functionals are used to design an output-feedback dynamic controller to satisfy the closed-loop stability and reference MAP tracking requirements. The synthesis conditions are formulated in terms of Linear Matrix Inequalities (LMIs) that characterize the induced $\mathcal{L}_2$-norm performance specification of the closed-loop system. The main objectives of the proposed control method in the presence of limitations posed by the time-varying model parameters and the large time-varying delay are to track the MAP reference command and maintain the blood pressure within the permissible range of commanded set-point, avoid undesirable overshoot and slow response, and to provide a smooth drug injection. Finally, to evaluate the performance of the proposed LPV blood pressure regulation approach, closed-loop simulations are conducted and the results confirm the effectiveness of the proposed control method against various simulated scenarios.}
\end{abstract}

\section*{INTRODUCTION}
\label{sec:intro}

The cardiovascular system can be seen as a hydraulic system that circulates blood as its working fluid and it is characterized in terms of pressure and flow variability. The mean arterial blood pressure (MAP) regulation of a patient to a desired target value is essential in many clinical and operative procedures in emergency resuscitation. For this purpose, two types of vasoactive drugs are being used to attain a target MAP in emergency resuscitation: (1) vasodilator drug to decrease the MAP to a target value, like sodium nitroprusside (SNP) which reduces the tension in the blood vessel walls \cite{he1986multiple}, (2) vasopressor drugs to increase the MAP to a target value, like phenylephrine (PHP) which stimulates the depressed cardiovascular system causing vasoconstriction \cite{neves2010phenylephrine}.

Typically, MAP control and regulation procedures in clinical care are carried out manually using a syringe or infusion pump with a manual titration by the medical personnel. In these cases, drug delivery and adjustment may not be precisely managed, which can lead to undesirable or potentially fatal consequences, such as, increased cardiac workload and cardiac arrest. Moreover, manual drug administration is a time-consuming and labor-intensive task and often is challenged by poor and sluggish performance. Further, inaccurate operator monitoring can lead to under- or over-resuscitation with potentially dangerous outcomes \cite{kee2005prevention}. Accordingly, the automation of vasoactive drug infusion via feedback control has been proposed as a potential remedy to tackle the mentioned challenges of manual drug administration.

Automated administration and accurate dosing of the intravenous drug is still a challenging problem mostly due to the pharmacological response variability of patients. There are considerable intra-patient and inter-patient variations in the physiological MAP response to the drug infusion \cite{kashihara2004adaptive}.  Moreover, the MAP response has a large varying time-delay, which poses significant limitations to the control design process.  Therefore, the designed control system should be robust against the complex physiological variations and disturbances and also be able to adapt to varying patient response and large varying time-delay. 

Time delay and especially large varying time delay presence considerably complicates the control design procedure and introduces instability and performance deterioration to the feedback system \cite{niculescu2012advances}. Stability analysis and control of time-delay systems has been the subject of research in recent decades and is classified into two general directions; namely,  delay-independent and delay-dependent approaches \cite{Fridman2014}. The delay-independent stabilization criteria are based on conditions that are independent of the size of the delay and the controller is designed regardless of the time delay magnitude, which leads to more conservative results especially for cases with small time delays (\textit{e.g.} see \cite{chen1994asymptotic}). On the other hand, the delay-dependent approaches result in less conservative analysis and guarantee the stability and the prescribed performance of the system by considering all the delays smaller than a certain determined upper bound. In recent years, there have been considerable efforts to investigate the stability of time-delay systems using delay-dependent stabilization criteria (\textit{e.g.} see \cite{loiseau2009topics, sakthivel2012robust, StasoujianRobust}). Delay-dependent approaches typically utilize Lyapunov-Krasovskii functionals \cite{kharitonov2004lyapunov} or Lyapunov-Razumnikhin functions \cite{jankovic2001control} for the stability analysis and control synthesis.  

The main challenge associated with the MAP regulation problem stems from the physiological and pharmacological variations of the considered system as well as its large varying time-delay, that restrict the application of conventional and time-invariant control techniques. Robust control methods treat the variations in the plant dynamics as uncertainties and employ time-invariant methods which leads to considerably conservative results \cite{schulz2016linear, dezfuli2016internal}. Early gain-scheduling control design approaches were based on interpolation, and despite their time-varying nature which adjusts for the system variations, they fail to provide any guarantee for asymptotic stability and/or performance of the time-varying closed-loop system. In this regard, the necessity to design gain-scheduled controllers with guaranteed stability and performance with less conservative results has led to the advent of linear parameter varying (LPV) gain-scheduling synthesis \cite{shamma1990analysis}.

LPV systems encompass linear systems whose dynamics depend on time-varying parameters known as scheduling parameters. For the present case study, in order to utilize the proposed output-feedback LPV control framework, an LPV time-delay system formulation is developed to describe the patient's MAP response to vasoactive drugs, where the system is a function of observable scheduling parameters. Stability analysis and control of LPV systems without time delay has been investigated in the literature \cite{apkarian1998advanced}. Time-delayed LPV systems where the time-varying delay is a function of the scheduling parameter are ubiquitous in engineering and the analysis and the control synthesis of those has been an interesting research topic in the past decade. One simple but conservative approach in dealing with time-delayed systems is using the Pad\'{e} approximation to transform the infinite-dimensional delay problem into a finite-dimensional non-minimum phase (NMP) system, followed by an LPV control design for a non-delay system \cite{Luspay2015, tasoujian2016parameter}. A common direct approach in analyzing LPV systems with time delay is to transfer the delay appearing in the input to the state. Such a system is called LPV state-delayed systems and has been addressed widely in the literature \cite{Luspay2015, zope2012delay}. In most of the control design methods for time-delayed LPV systems, Lyapunov-Krasovskii functionals are considered (see, \textit{e.g.} \cite{zope2012delay}). In these studies, the delay variation rate is not taken into account, which resulted in conservative results. 

In the present paper, we utilize parameter-dependent Lyapunov-Krasovskii functionals to examine the stability and the tracking performance of the closed-loop system, which leads to a delay-dependent synthesis method that can handle arbitrary parameter variations. Accordingly, an LPV output-feedback controller is designed to track a desired MAP reference and minimize the effect of disturbances. For control validation purposes, collected animal experiment data and a non-linear patient simulation model are utilized to capture the varying physiological characteristics of a patient's MAP response \cite{Luspay2015}. The proposed approach is evaluated in computer simulations and compared to the results of a conventional fixed gain proportional integral (PI) controller which has been validated experimentally (see \cite{wassar2014automatic}).

The notation to be used in the paper is standard and as follows: $\mathbb{R}$ denotes the set of real numbers, $\mathbb{R}_{+}$ is the set of non-negative real numbers, $\mathbb{R}^n$ and $\mathbb{R}^{k \times m}$ are used to denote the set of real vectors of dimension $n$ and the set of real $k \times m$ matrices, respectively. $M \succ \mathbf{0}$ shows the positive definiteness of the matrix $M$  and the transpose of a real matrix $M$ is shown as $M^{\text{T}}$ . Also, $\mathbb{S}^{n}$ and $\mathbb{S}^{n}_{++}$ denote the set of real symmetric and real symmetric positive definite $n \times n$ matrices, respectively. In a symmetric matrix, the asterisk $\star$ in the $(i,\: j)$ element denotes transpose of $(j,\: i)$ element. $\mathscr{C} (J,\: K)$ denotes the set of continuous functions from a set $J$ to a set $K$.

The outline of the paper is as follows. Section 3 introduces a simplified mathematical representation of the studied MAP response dynamics followed by the control design objectives. In section 4, the LPV model of patients' MAP response is presented and LPV gain-scheduling control design is introduced. Section 5 will outline closed-loop results and evaluate the performance of the proposed LPV controller in the computer simulation environment. Finally, section 6 will conclude the paper. 

\section*{MAP DRUG RESPONSE MODEL}
\label{sec:ProbForm}

\begin{figure}[t] 
\centering \includegraphics[width=3.6in, height=2.3in]{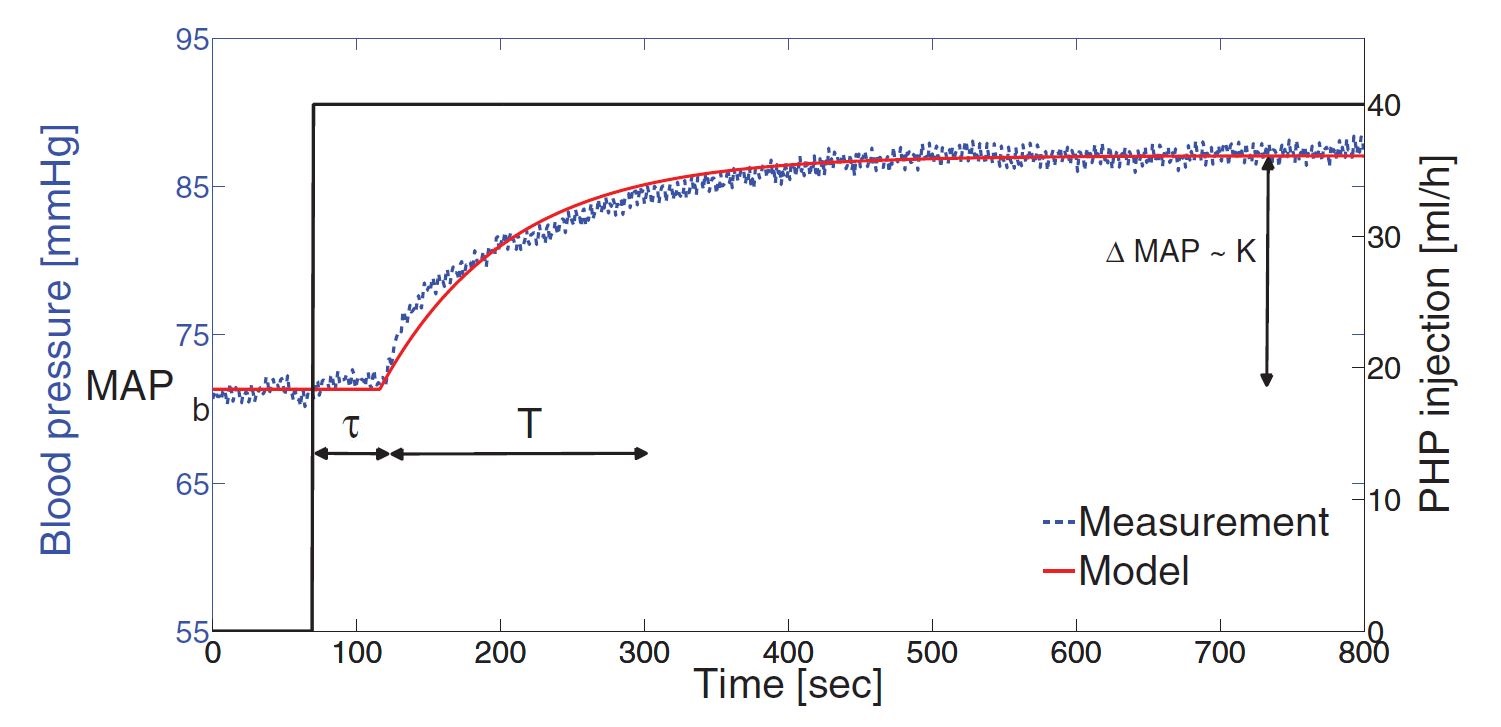} 
\caption{Typical MAP response due to step PHP drug infusion} 
\label{fig:MAPresponse}
\end{figure}

In the present study, we consider a first-order model with a time-delay to describe the patient's MAP response to the infusion of a vasoactive drug, such as PHP, \textit{i.e.}

\begin{equation}
G_p(s)=\frac{\Delta MAP (s)}{U(s)}=\frac{K}{T s+1}e^{-\tau s},
\label{eq:MAP response TF}
\end{equation}

\noindent where $\Delta MAP$ stands for the MAP variations in \textit{mmHg} from its baseline value,  \textit{i.e.} $\Delta MAP (t)= MAP (t) -MAP_b$, $U$  is the drug infusion rate in \textit{ml/h}, $K$ denotes the patient's sensitivity to the drug, $T$ is the lag time representing the uptake, distribution, and biotransformation of the drug \cite{isaka1993control}, and $\tau$ is the time delay for the drug to reach the circulatory system from the infusion pump. This first-order model seems to properly capture a patient's physiological response to the PHP drug injection. Figure~\ref{fig:MAPresponse} shows a typical MAP response to the step PHP infusion versus a matched response of the considered model (\ref{eq:MAP response TF}). Data is collected from swine experiments performed at the Resuscitation Research Laboratory of the University of Texas Medical Branch (UTMB), Galveston, Texas \cite{StasoujianRobust}. Although the proposed model structure (\ref{eq:MAP response TF}) is qualitatively able to represent the characteristics of the MAP response to the infusion of PHP, the model parameters vary considerably over time due to the variability of patients' pharmacological response to the vasoactive drug infusion. That is, the model parameters and delay could vary significantly from patient-to-patient (inter-patient variability), as well as, for a given patient over time (intra-patient variability) \cite{isaka1993control, rao2003experimental}
. Thus, these variations cause the model parameters $K$, $T$, and $\tau$ to be time-varying and \textit{a priori} unknown.


The designed controller should be able to guarantee stability and tracking performance despite these variations. 
The main purpose of this work is to design such a controller to automatically regulate the blood pressure to a target value and maintain hemodynamic stability in hypotensive patients in the presence of large time-varying delay and model parameter variability by means of the closed-loop administration of the vasopressor drug, PHP. 
The following section, firstly, presents an LPV model for the patient's MAP response and then, a delay-dependent output-feedback control strategy is designed to satisfy the prescribed closed-loop stability and reference tracking performance.


\section*{LPV MODELING AND LPV GAIN-SCHEDULING CONTROL DESIGN}
\label{sec:num3}

\begin{figure}[t]
	\centering
		\begin{tikzpicture}[auto, font = {\sf \scriptsize}, cross/.style={path picture={\draw[black] (path picture bounding box.south east) -- (path picture bounding box.north west) (path picture bounding box.south west)-- (path picture bounding box.north east);}}, node distance=2cm,>=latex]
		\node at (-5,0)[input, name=input1] {};
		\node at (-2.2,0)[block,text width=1.7cm,text height=-.2em,text depth=0em] (Controller) {{\centering LPV  \\ Controller}};
 	    \node at (.6,0)[block,text width=.8cm,text height=.7em,text depth=.2em] (Patient) {\centering Patient};
 	    \node at (.6,1.5)[block,text width=1.2cm,text height=-.3em,text depth=.1em] (ParamEstim) {\centering Parameter \\ Generation};
		\node at (2,.7)[input, name=Disturb] {}; 	    
	    \node at (-4,0) [sum,cross] (sum1) {};
	    \node at (2,0) [sum,cross] (sum2) {};
	    
	    \node [output, left of=Patient, node distance=1.1cm] (output1) {};
	    \node [output, right of=Patient, node distance=1.1cm] (output2) {};
	    \node [output, right of=sum2, node distance=1.25cm] (output3) {};
	    \node [output, right of=sum2, node distance=.6cm] (output4) {};

	    \draw [->] (input1) -- node[name=in1tosum1, pos=0.2] {$y^*(t)$}node[name=ms2, pos=.8, below] {$+$} (sum1);
		\draw [->] (sum1) -- node[name=sum1tosum2, pos=.4] {$e(t)$} (Controller);	 	    
	    \draw [->] (Controller) -- node[name=sum2toCont, pos=0.3] {$u(t)$} (Patient);
	    \draw [->] (Patient) -- node[name=sum2toCont, pos=0.8] {$+$} (sum2); 
	    \draw [->] (sum2) -- node[name=sum2toCont, pos=0.9, above] {$y$} (output3); 
	    \draw [->] (output4) -- +(0,-1) -| node[name=sum2toCont, pos=0.9, right] {$-$} (sum1);
	    \draw [->] (output4) -- +(0,1.5) -- node[name=sum2toCont, pos=0.9, right] {} (ParamEstim.360);
	    
	    \draw [->] (output1) |- node[name=sum2toCont, pos=0.9] {} (ParamEstim.195);

	    \draw [->] (ParamEstim.160) -| node[name=sum5tosum1, pos=0.9, left] {$K$} (Controller.145); 
	    \draw [->] (ParamEstim.170) -| node[name=sum5tosum1, pos=0.9] {$T$} (Controller.115); 
	    \draw [->] (ParamEstim.180) -| node[name=sum5tosum1, pos=0.9] {$\tau$} (Controller.50); 	    
	    \draw [->] (Disturb) -- node[pos=.1, above]{$d(t)$} node[name=sum5tosum1, pos=0.9] {$+$} (sum2); 	    
	\end{tikzpicture}
    \caption{Closed-loop system structure}
    \label{fig:system structure}
\end{figure}
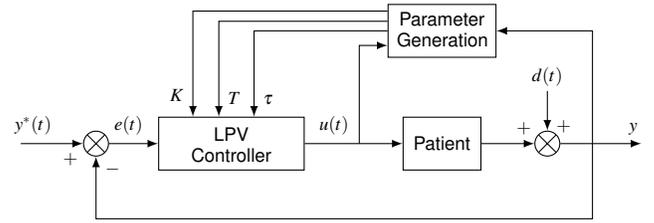

In order to apply the linear parameter varying (LPV) control strategy to the MAP regulation problem, the MAP response model (\ref{eq:MAP response TF}) is formulated as an LPV time-delay model, where the model parameters are continuously available for the controller as scheduling parameters. The structure of the closed-loop system with the LPV controller is shown in Fig.~\ref{fig:system structure}. 

\subsection*{MAP Drug Response LPV Modeling}

We consider  (\ref{eq:MAP response TF}) as a first-order model with time-delay to capture patients' MAP response dynamics to the PHP infusion and its equivalent time-domain representation as
\begin{equation} 
\label{eq:MAP response time domain}
T \cdot \dot{\Delta MAP (t)} + \Delta MAP (t) = K \cdot u(t-\tau).
\end{equation}
By considering the state variable as $x(t) = \Delta MAP (t)$, we can rewrite the state space representation of the first-order time-delayed MAP response model as follows

\begin{equation}
\begin{matrix}
 \dot{x}(t) & = & -\dfrac{1}{T} x(t) + \dfrac{K}{T} u(t-\tau),\\[0.2cm]
 y(t)& = & x(t) + d_o(t),\:\:\:\:\:\:\:\:\:\:\:
\end{matrix}
\label{eq:MAP state space}
\end{equation}
\noindent where $y(t)$ is the patient's measured MAP response and $d_o(t)$ denotes output disturbance. In (\ref{eq:MAP state space}), the varying time delay $\tau = \tau(t)$ is appearing in the input signal and in order to utilize the proposed time-delay LPV control design framework, we need to transform the input-delay system into a state-delay LPV representation. To this end, we introduce a filtered input signal $u_a(t)$ as follows
\begin{equation}
u(s)=\frac{\Omega}{s + \Lambda} u_a (s),
\end{equation}
\noindent where $\Omega$ and $\Lambda$ are positive scalars that are selected based on the bandwidth of the actuators. By considering the augmented state vector  $\mathbf{x}^{\text{T}} _a = [\begin{array}{ccc}
	x & u & x_e\end{array}]^{\text{T}}$, and defining the scheduling parameter vector, $\boldsymbol{\rho}(t) = [\begin{array}{ccc}
	K(t) & T(t) & \tau(t) \end{array}]^{\text{T}}$, the LPV state-delayed state-space representation of the MAP response dynamics takes the following form
\begin{equation}
\begin{array}{rl}
 \dot{\mathbf{x}}_a(t)& =  \mathbf{A}(\boldsymbol{\rho}(t)) \mathbf{x}_a(t)+\mathbf{A}_d(\boldsymbol{\rho}(t)) \mathbf{x}_a(t-\tau(t)) + \mathbf{B}_1(\boldsymbol{\rho}(t))\mathbf{w}(t)   \\ 
& + \mathbf{B}_2(\boldsymbol{\rho}(t))u(t) \\[0.20cm] 
 y_a(t) & = \mathbf{C}_2(\boldsymbol{\rho}(t)) \mathbf{x}_a(t)+ \mathbf{C}_{2d}(\boldsymbol{\rho}(t)) \mathbf{x}_a(t-\tau(t))\\
 & + \mathbf{D}_{21}(\boldsymbol{\rho}(t)) \mathbf{w}(t),
\end{array}
\label{LPV_MAP}
\end{equation}
where the exogenous disturbance vector $\mathbf{w}(t) = [\begin{array}{cc}
 																												r(t) & d_o(t)\end{array} ]^{\text{T}}$ includes the reference command and output disturbance. The third state is defined for the command tracking purposes, \textit{i.e.} $\dot{x}_e (t) = e(t) = r(t)-y(t) = r(t) - (x(t) + d_o(t))$. Thus, the state space matrices of the augmented LPV system (\ref{LPV_MAP}) are obtained as

\begin{align}
			& \mathbf{A}(\boldsymbol{\rho}(t)) =\begin{bmatrix}
- \frac{1}{T(t)} & 0 & 0\\ 
0 & -\Lambda & 0 \\ 
-1 & 0 & 0
\end{bmatrix}, \: \mathbf{A}_d(\boldsymbol{\rho}(t))=\begin{bmatrix}
0 & \frac{K(t)}{T(t)} & 0\\ 
0 & 0 & 0 \\ 
0 & 0 & 0
\end{bmatrix},\nonumber\\
		& \mathbf{B}_1(\boldsymbol{\rho}(t)) =\begin{bmatrix}
0 & 0\\ 
0 & 0\\ 
1 & -1
\end{bmatrix}, \mathbf{B}_2(\boldsymbol{\rho}(t))=\begin{bmatrix}
0\\ 
\Omega\\ 
0
\end{bmatrix},\nonumber\\
			& \mathbf{C}_2(\boldsymbol{\rho}(t)) = \begin{bmatrix}
1 & 0 & 0 \end{bmatrix}, \: \mathbf{D}_{21}(\boldsymbol{\rho}(t))=\begin{bmatrix}
0 & 1
\end{bmatrix},
\label{eq:matrices1}
\end{align}
and $\mathbf{C}_{2d}(\boldsymbol{\rho}(t))$ is a zero matrix with compatible dimensions. Now, by having an appropriate LPV representation of the considered MAP response model, the proposed LPV gain-scheduling synthesis is discussed in the following section.

\subsection*{LPV Time-Delay Control Design}
\label{sec:control design}

Consider the following state-space representation of an LPV system with a varying state delay

\begin{equation}
\begin{array}{cl}
 \dot{\mathbf{x}}(t) & =  \mathbf{A}(\boldsymbol{\rho}(t)) \mathbf{x}(t)+\mathbf{A}_d(\boldsymbol{\rho}(t)) \mathbf{x}(t-\tau(t)) + \mathbf{B}_1(\boldsymbol{\rho}(t))\mathbf{w}(t)\\
 & + \mathbf{B}_2(\boldsymbol{\rho}(t))\mathbf{u}(t) \\[0.25cm] 
  \mathbf{z}(t) & = \mathbf{C}_1(\boldsymbol{\rho}(t)) \mathbf{x}(t) + \mathbf{C}_{1d}(\boldsymbol{\rho}(t)) \mathbf{x}(t-\tau(t)) + \mathbf{D}_{11}(\boldsymbol{\rho}(t)) \mathbf{w}(t)\\
  & + \mathbf{D}_{12}(\boldsymbol{\rho}(t)) \mathbf{u}(t)\\[0.25cm]
 \mathbf{y}(t)& = \mathbf{C}_2(\boldsymbol{\rho}(t)) \mathbf{x}(t)+ \mathbf{C}_{2d}(\boldsymbol{\rho}(t)) \mathbf{x}(t-\tau(t)) + \mathbf{D}_{21}(\boldsymbol{\rho}(t)) \mathbf{w}(t),\\[0.25cm]
 \mathbf{x}(\theta)  & = \boldsymbol{\phi}(\theta),\forall \theta \in [-\tau(\boldsymbol{\rho}(0)), \: \: 0],
\end{array}
\label{LPVsystem}
\end{equation}

\noindent where $\mathbf{x}(t) \in \mathbb{R}^n$ is the system state vector, $\mathbf{w}(t) \in \mathbb{R}^{n_w}$ is the vector of exogenous disturbances with finite energy in the space $\mathcal{L}_2[0, \:\: \infty]$, $\mathbf{u}(t) \in \mathbb{R}^{n_u}$ is the input vector, $\mathbf{z}(t) \in \mathbb{R}^{n_z}$ is the vector of outputs to be controlled, $\mathbf{y}(t) \in \mathbb{R}^{n_y}$ is the vector of measurable outputs, $\boldsymbol{\phi}(\cdot)$ is the initial system condition, and $\tau$ is a differentiable scalar function representing the parameter-varying time delay. In the present work, the delay is assumed bounded and the function $\tau$ lies in the set $\mathscr{T}$ defined as

\begin{equation}
\mathscr{\mathscr{T}} := \{ \tau \in \mathscr{C}(\mathbb{R}^{s},\mathbb{R}) : 0 \leq \tau(t) \leq \overline{\tau} < \infty, \forall t \in \mathbb{R}_{+}\}.
\end{equation}
Also, the system's initial condition function, $\boldsymbol{\phi}$ is a given function in $\mathscr{C}([\begin{array}{cc}
-\overline{\tau} & 0\end{array}], \mathbb{R}^n)$. The notation $\mathbf{x}_t(\theta)$ refers to $\mathbf{x}(t+\theta)$ for $\theta \in [\begin{array}{cc}
-\overline{\tau} & 0\end{array}]$ where $\mathbf{x}_t$ is the infinite-dimensional state vector of the system. The presented state space matrices $\mathbf{A}(\cdot)$, $\mathbf{A}_d(\cdot)$, $\mathbf{B}_1(\cdot)$, $\mathbf{B}_2(\cdot)$, $\mathbf{C}_1(\cdot)$, $\mathbf{C}_{1d}(\cdot)$, $\mathbf{C}_2(\cdot)$, $\mathbf{C}_{2d}(\cdot)$, $\mathbf{D}_{11}(\cdot)$, $\mathbf{D}_{12}(\cdot)$, and $\mathbf{D}_{21}(\cdot)$ are known continuous functions of the time-varying parameter vector $\boldsymbol{\rho}(\cdot) \in \mathscr{F}^\nu _\mathscr{P}$, where $\mathscr{F}^\nu _\mathscr{P}$ is the set of allowable parameter trajectories defined as

\begin{equation}
\mathscr{F}^\nu _\mathscr{P} \triangleq \{\boldsymbol{\rho} \in \mathscr{C}(\mathbb{R}_{+},\mathbb{R}^{s}):\boldsymbol{\rho}(t) \in \mathscr{P}, |\dot{\rho}_i (t)| \leq \nu_i \},
\end{equation}

\noindent wherein  $\mathscr{P}$ is a compact subset of $\mathbb{R}^{s}$, \textit{i.e.}, the parameter trajectories and parameter variation rates are assumed bounded as defined. Since the delay is considered to be dependent on the scheduling parameter vector $\boldsymbol{\rho}(t)$, as a result, the delay bound should be incorporated into the parameter set $\mathscr{F}^\nu _\mathscr{P}$. Moreover, in the present paper, Lyapunov Krasovskii functionals are utilized to obtain less conservative results, which are valid for bounded parameter variation rates \cite{apkarian1998advanced}.

In the present work, we use an $\mathcal{H}_{\infty}$ design performance specification for the closed-loop system. The induced $\mathcal{L}_2$-gain (or $\mathcal{H}_{\infty}$ norm) of the LPV system (\ref{LPVsystem}) from $\mathbf{w}(t)$ to $\mathbf{z}(t)$ is defined as

\begin{equation}
\Vert \mathbf{T}_{\mathbf{z}\mathbf{w}}\Vert_{i,2} = \underset{\boldsymbol{\rho}(t) \in \mathscr{F}^\nu _\mathscr{P}}{\sup} \:\:\: \underset{\Vert \mathbf{w}(t) \Vert_2 \neq 0}{\sup}\:\: \frac{\Vert \mathbf{z}(t) \Vert_2}{\Vert \mathbf{w}(t) \Vert_2}.
				\label{eq:Performance Index}
\end{equation}

\noindent where $\mathbf{T}_{\mathbf{z}\mathbf{w}}$ is an operator mapping $\mathbf{w}(t)$ to $\mathbf{z}(t)$, and $\Vert \mathbf{w}(t) \Vert_2$ and $\Vert \mathbf{z}(t) \Vert_2$ are the 2-norm of the exogenous input signal and desired controlled output vector, respectively.

Considering the time-delay LPV system (\ref{LPVsystem}) and $\mathbf{u}(t)\equiv0$, the following theorem provides the sufficient condition that guarantees the asymptotic stability and the specified level of disturbance rejection performance in the  $\mathcal{H}_{\infty}$  setting as defined in (\ref{eq:Performance Index}).

\begin{theorem}\label{thm:thm1}\cite{zope2012delay} The unforced time-delay LPV system (\ref{LPVsystem}) is asymptotically stable for all $\tau \in \mathscr{T}$ and satisfies the condition $\Vert \mathbf{z}(t) \Vert_2 \leq \gamma \Vert \mathbf{w}(t) \Vert_2$, if there exists a continuously differentiable matrix function $\mathbf{P}:\mathbb{R}^{s}\rightarrow\mathbb{S}^{n}_{++}$, constant matrices $\mathbf{Q}$, $\mathbf{R} \in \mathbb{S}^{n}_{++}$, and a scalar $\mathbf{\gamma} > 0$ such that the following LMI condition holds for all $\boldsymbol{\rho} \in \mathscr{F}^\nu _\mathscr{P}$.

\begin{equation}
\left[\begin{array}{ccccc}
(1,1) & \mathbf{P}\mathbf{A}_d + \mathbf{R} &\:\:\:\: \mathbf{P} \mathbf{B}_1 & \:\:\mathbf{C}_1 ^\text{T} & \:\:\:\: 
 \overline{\tau} A^\text{T} \mathbf{R}\\  [1mm]
 \star & \boldsymbol{\Xi}_{22} & \:\:\:\: \mathbf{0} &\:\:\:\: \mathbf{C}_{1d}^\text{T} & \:\:\:\:  \overline{\tau} \mathbf{A}_{d}^\text{T}\mathbf{R}\\ [1mm]
\star & \star  & -\gamma\mathbf{I} & \:\:\:\: \mathbf{D}_{11}^ \text{T} & \:\:\:\:  \overline{\tau} \mathbf{B}_1^\text{T}\mathbf{R}\\ [1mm]
\star & \star & \star &  -\gamma\mathbf{I} &\:\:\:\:\:\:\:\:  \mathbf{0}\\ [1mm]
\star & \star & \star  & \star &\:\:\:\:  -\mathbf{R} \\
\end{array}\right]\prec\mathbf{0},
\label{hinf performance lmi condition 1}
\end{equation}

\noindent where $(1,1)=\mathbf{A}^{\text{T}}\mathbf{P}+ \mathbf{P}\mathbf{A}+ \dfrac{\partial \mathbf{P}}{\partial \rho_i}\dot{\rho}_i  + \mathbf{Q}-\mathbf{R}$, and $\boldsymbol{\Xi}_{22}=- \big(1-   \dfrac{\partial \tau}{\partial \rho_i} \dot{\rho}_i \big)  \mathbf{Q}-\mathbf{R}$ and and the dependence on the scheduling parameter has been dropped for the clarity and simplicity. It should be noted that the parameter variation rate $\dot{\rho}$, enters affinely in the LMIs in this paper, hence, it is enough to check LMI only at the vertices of $\dot{\rho}$. As a result, $\dfrac{\partial \mathbf{P}}{\partial \rho_i} \dot{\rho}_i $ and $\dfrac{\partial \tau}{\partial \rho_i} \dot{\rho}_i $ are replaced by $\sum_{i=1}^s \pm \bigg(\nu_i \dfrac{\partial \mathbf{P}}{\partial \rho_i}\bigg)$  and $\sum_{i=1}^s \pm \bigg(\nu_i \dfrac{\partial \tau}{\partial \rho_i} \bigg)$, respectively.

\end{theorem}

The standard matrix inequality characterization presented in Theorem \ref{thm:thm1} has multiple product terms of $\mathbf{P} \mathbf{A}$ and $\mathbf{R} \mathbf{A}$, so it is not suitable to derive the synthesis conditions in which the closed-loop $\mathbf{A}$ matrix includes the unknown terms of the controller to be designed. To overcome this challenge, a reciprocal variant of projection Lemma \cite{zope2012delay} is utilized to derive a relaxed condition. By using this approach, slack variables are introduced which bring more flexibility in the synthesis problem and lead to less conservative conditions. The following theorem presents a relaxed LMI condition using slack variables to guarantee the asymptotic stability and the specified level of performance in the $\mathcal{H}_{\infty}$  setting given in  (\ref{eq:Performance Index}).

\begin{theorem}\label{thm:thm2}\cite{zope2012delay} The time-delay LPV system (\ref{LPVsystem}) when $\mathbf{u}\equiv0$ is asymptotically stable for all $\tau \in \mathscr{T}$ and satisfies the condition $\Vert \mathbf{z}(t) \Vert_2 \leq \gamma \Vert \mathbf{w}(t) \Vert_2$, if there exists a continuously differentiable matrix function $\mathbf{P}:\mathbb{R}^{s}\rightarrow\mathbb{S}^{n}_{++}$, constant matrices $\mathbf{Q}$, $\mathbf{R} \in \mathbb{S}^{n}_{++}$, matrix functions $\mathbf{V}_1$, $\mathbf{V}_2$, $\mathbf{V}_3$ : $\mathbb{R}^{s}\rightarrow\mathbb{R}^{n \times n}$, and a scalar $\mathbf{\gamma} > 0$ such that the following LMI condition holds for all $\boldsymbol{\rho} \in \mathscr{F}^\nu _\mathscr{P}$.
\begin{small}
\begin{equation}
\begin{array}{l}
\left[\begin{array}{ccc}
-\mathbf{V}_1 - \mathbf{V}^{\text{T}}_1 & \mathbf{P}- \mathbf{V}^{\text{T}}_2 + \mathbf{V}_1 \mathbf{A}  &- \mathbf{V}^{\text{T}}_3 + \mathbf{V}_1 \mathbf{A}_d \\ [1mm]
 \star & \boldsymbol{\Psi}_{22}+\mathbf{A}^{\text{T}}\mathbf{V}^{\text{T}}_2+\mathbf{V}_2\mathbf{A}  & \mathbf{R}+\mathbf{A}^{\text{T}}\mathbf{V}^{\text{T}}_3+\mathbf{V}_2 \mathbf{A}_d \\ [1mm]
\star & \star & \boldsymbol{\Xi}_{22}+\mathbf{A}_d^{\text{T}}\mathbf{V}^{\text{T}}_3+\mathbf{V}_3 \mathbf{A}_d\\ [1mm]
\star & \star & \star\\
\star & \star & \star\\
\star & \star & \star 
\end{array}\right.\\ [2mm]
\qquad\qquad\qquad\qquad\qquad\qquad\quad\quad\quad\left.\begin{array}{ccc}
\mathbf{V}_1 \mathbf{B}_1& \:\:\: \mathbf{0} & \mathbf{V}_1+\overline{\tau}\mathbf{R} \\ [1mm]
\mathbf{V}_2 \mathbf{B}_1 & \:\:\: \mathbf{C}^{\text{T}}_1 & \mathbf{V}_2 - \mathbf{P}\\ [1mm]
\mathbf{V}_3 \mathbf{B}_1 & \:\:\: \mathbf{C}^{\text{T}}_{1d} & \mathbf{V}_3\\  [1mm]
-\gamma \mathbf{I} & \:\:\: \mathbf{D}^{\text{T}}_{11} & \mathbf{0}\\ [1mm]
 \star & \:\:\: -\gamma \mathbf{I} & \mathbf{0}\\ [1mm]
\star & \:\:\: \star  & (-1-2\overline{\tau})\mathbf{R} 
\end{array}\right]\prec\mathbf{0},
\end{array}
\label{hinf performance lmi condition slack variables}
\end{equation}
\end{small}
\noindent where $\boldsymbol{\Psi}_{22} = \dfrac{\partial \mathbf{P}}{\partial \rho_i}\dot{\rho}_i + \mathbf{Q} - \mathbf{R}$ and $\boldsymbol{\Xi}_{22}$ has been defined before.

\end{theorem}

The analysis results presented in Theorem \ref{thm:thm2} are used to derive a dynamic output feedback controller. In this paper, the time delay in the system dynamics has been considered to be known and measurable function of the scheduling parameter $\boldsymbol{\rho}$. A full-order dynamic output-feedback controller in the following form is designed for the time-delay LPV system (\ref{LPVsystem}):
\begin{equation}
\begin{array}{cl}
\dot{\mathbf{x}}_k(t) & = \mathbf{A}_k (\boldsymbol{\rho}) \mathbf{x}_k(t)+ \mathbf{A}_{dk}(\rho)\mathbf{x}_k(t-\tau(t))+\mathbf{B}_k(\boldsymbol{\rho})\mathbf{y}(t),\\ 
\mathbf{u}(t) & = \mathbf{C}_k (\boldsymbol{\rho}) \mathbf{x}_k(t)+ \mathbf{C}_{dk}(\boldsymbol{\rho})\mathbf{x}_k(t-\tau(t))+\mathbf{D}_k(\boldsymbol{\rho})\mathbf{y}(t),
\end{array}
\label{controller}
\end{equation}
where $\mathbf{x}_k (t) \in \mathbb{R}^{n}$ is the controller state vector and $\mathbf{x}_k(t-\tau(t)) \in \mathbb{R}^{n}$ is the delayed state of the controller. Considering the system dynamics (\ref{LPVsystem}) and the controller (\ref{controller}) the closed-loop system would be as follows
\begin{equation}
\begin{array}{cl}
\dot{\mathbf{x}}_{cl}(t) & =  \mathbf{A}_{cl}\: {\mathbf{x}}_{cl}(t) + \mathbf{A}_{d,cl}\: {\mathbf{x}}_{cl}(t-\tau(t)) + \mathbf{B}_{cl}\:\mathbf{w}(t),\\ 
\mathbf{z}(t) & =  \mathbf{C}_{cl}\:{\mathbf{x}}_{cl}(t) +  \mathbf{C}_{d,cl}\:{\mathbf{x}}_{cl}(t-\tau(t)) + \mathbf{D}_{cl}\: \mathbf{w}(t),
\end{array}
\label{eq:closed-loop system}
\end{equation}
where 
\begin{align*}
		& \mathbf{A}_{cl}=\begin{bmatrix}
\mathbf{A} + \mathbf{B}_2 \mathbf{D}_k \mathbf{C}_2 & \mathbf{B}_2 \mathbf{C}_k\\ 
\mathbf{B}_k \mathbf{C}_2 & \mathbf{A}_k 
\end{bmatrix}, \\
& \mathbf{A}_{d,cl}\!=\!\begin{bmatrix}
\mathbf{A}_d + \mathbf{B}_2 \mathbf{D}_k \mathbf{C}_{2d} & \mathbf{B}_2 \mathbf{C}_{dk}\\ 
\mathbf{B}_k \mathbf{C}_{2d} & \mathbf{A}_{dk} 
\end{bmatrix},  \mathbf{B}_{cl}=\begin{bmatrix}
\mathbf{B}_1 + \mathbf{B}_2 \mathbf{D}_k \mathbf{D}_{21} \\
\mathbf{B}_k \mathbf{D}_{21} 
\end{bmatrix}, \\
& \mathbf{C}_{cl}=\begin{bmatrix}
\mathbf{C}_1 + \mathbf{D}_{12} \mathbf{D}_k \mathbf{C}_2 & \mathbf{D}_{12} \mathbf{C}_k
\end{bmatrix}, \\
& \mathbf{C}_{d,cl} =\begin{bmatrix}
\mathbf{C}_{1d} + \mathbf{D}_{12} \mathbf{D}_k \mathbf{C}_{2d} & \mathbf{D}_{12} \mathbf{C}_{dk}
\end{bmatrix}, \\
& \mathbf{D}_{cl}= \mathbf{D}_{11} + \mathbf{D}_{12} \mathbf{D}_k \mathbf{D}_{21}, 
\end{align*}
\noindent and $\mathbf{x}_{cl}(t) = [\begin{array}{cc}
\mathbf{x}(t) &\mathbf{x}_k (t)\end{array} ]^{\text{T}}$, and the dependence on the scheduling parameter has been dropped for clarity. Now, considering the closed-loop system (\ref{eq:closed-loop system}), the following result provides sufficient conditions for the synthesis of a delayed output-feedback controller which guarantees closed-loop asymptotic stability and a specified level of performance as defined earlier in (\ref{eq:Performance Index}).

\begin{theorem}\label{thm:thm3}\cite{Briat2015} The system (\ref{LPVsystem}) is asymptotically stable for parameters $\boldsymbol{\rho} \in \mathscr{F}^\nu _\mathscr{P}$ and all delays $\tau \in \mathscr{T}$ and satisfy the condition $||\mathbf{z}||_2 \leq \gamma ||\mathbf{w}||_2$ for the closed-loop system (\ref{eq:closed-loop system}), if there exists a continuously differentiable matrix function $\widetilde{\mathbf{P}} : \mathbb{R}^{s}\rightarrow\mathbb{S}^{2n}_{++}$, parameter dependent matrix functions $\mathbf{X}, \mathbf{Y}  : \mathbb{R}^{s}\rightarrow\mathbb{S}^{n}_{++}$, constant matrices $\widetilde{\mathbf{Q}}$, $\widetilde{\mathbf{R}} \in \mathbb{S}^{n}_{++}$, parameter dependent matrices $\widehat{A}$, $\widehat{A}_d$, $\widehat{B}$, $\widehat{C}$, $\widehat{C}_d$, $\widehat{D}_k$, three scalars $\mathbf{\gamma} > 0$, and $\lambda_2$, $\lambda_3$  such that the following LMI condition holds
\begin{equation}
\begin{array}{l}
\left[\begin{array}{ccc}
-2\widetilde{\mathbf{V}} & \widetilde{\mathbf{P}}-\lambda_2\widetilde{\mathbf{V}} + \mathscr{A} & -\lambda_3\widetilde{\mathbf{V}} + \mathscr{A}_d\\
\star & \widetilde{\boldsymbol{\Psi}}_{22} + \lambda_2(\mathscr{A}+\mathscr{A}^{\text{T}}) &  \widetilde{\mathbf{R}} + \lambda_3 \mathscr{A}^{\text{T}} + \lambda_2 \mathscr{A}_d \\
\star & \star & \widetilde{\boldsymbol{\Xi}}_{22} + \lambda_3(\mathscr{A}_d+\mathscr{A}_d^{\text{T}}) \\
\star & \star & \star \\
\star & \star & \star \\
\star & \star & \star  
\end{array}\right.\\
\qquad\qquad\qquad\qquad\quad\quad\quad\qquad\left.\begin{array}{ccc}
\mathscr{B} & 0 & \widetilde{\mathbf{V}}+\overline{\tau}\widetilde{\mathbf{R}} \\
\lambda_2 \mathscr{B}  &  \mathscr{C}^{\text{T}} & \lambda_2\widetilde{\mathbf{V}} - \widetilde{\mathbf{P}}\\
\lambda_3 \mathscr{B} & \mathscr{C}_d^{\text{T}}  & \lambda_3\widetilde{\mathbf{V}}\\ 
-\gamma \mathbf{I} & \mathscr{D}^{\text{T}} & 0\\
\star & -\gamma \mathbf{I} & 0\\
\star & \star & (-1 -2\overline{\tau})\widetilde{\mathbf{R}} 
\end{array}\right]\prec\mathbf{0},
\end{array}
\label{eq:LMI closed-loop}
\end{equation}
with
\begin{equation}
\begin{array}{lll}
\widetilde{\mathbf{V}}  & = & \begin{bmatrix}
\mathbf{Y} & \mathbf{I}\\ 
\mathbf{I} & \mathbf{X}
\end{bmatrix},\\[0.3cm]
\mathscr{A} & = &\begin{bmatrix}
\mathbf{A} \mathbf{Y}+\mathbf{B}_2\widehat{C} & \:\: \mathbf{A} +\mathbf{B}_2 \mathbf{D}_k \mathbf{C}_2\\ 
\widehat{A} & \:\: \mathbf{X} \mathbf{A} +\widehat{B}\mathbf{C}_2
\end{bmatrix} ,\\[0.4cm]
\mathscr{A}_d & = & \begin{bmatrix}
\mathbf{A}_d\mathbf{Y}+\mathbf{B}_2\widehat{C}_d & \:\: \mathbf{A}_d+\mathbf{B}_2 \mathbf{D}_k \mathbf{C}_{2d}\\ 
\widehat{A}_d & \:\: \mathbf{X}\mathbf{A}_d+\widehat{B}\mathbf{C}_{2d}
\end{bmatrix},\\[0.4cm]
\mathscr{B} & = & \begin{bmatrix}
\mathbf{B}_1 + \mathbf{B}_2 \mathbf{D}_k \mathbf{D}_{21}\\ 
\mathbf{X}\mathbf{B}_1 + \widehat{B} \mathbf{D}_{21}
\end{bmatrix},\\[0.3cm] 
\mathscr{C} & = & \begin{bmatrix}
\mathbf{C}_1 \mathbf{Y}+ \mathbf{D}_{12} \widehat{C} & \:\:\:\: \mathbf{C}_1 + \mathbf{D}_{12} \mathbf{D}_k \mathbf{C}_2 
\end{bmatrix},\\[0.3cm]
\mathscr{C}_d & = & \begin{bmatrix}
\mathbf{C}_{1d} \mathbf{Y}+ \mathbf{D}_{12} \widehat{C}_d & \:\:\:\: \mathbf{C}_{1d} + \mathbf{D}_{12} \mathbf{D}_k \mathbf{C}_{2d} 
\end{bmatrix},\\[0.3cm]
\mathscr{D} & = & \begin{bmatrix}
\mathbf{D}_{11} + \mathbf{D}_{12} \mathbf{D}_k \mathbf{D}_{21}
\end{bmatrix},\\[0.3cm] \widetilde{\boldsymbol{\Psi}}_{22} & = & \bigg[ \sum_{i=1}^s \pm \Big(\nu_i \frac{\partial \widetilde{\mathbf{P}}(\boldsymbol{\rho})}{\partial \rho_i}\Big) \bigg] + \widetilde{\mathbf{Q}}-\widetilde{\mathbf{R}}  ,\\[0.3cm]
\widetilde{\boldsymbol{\Xi}}_{22} & = &-\bigg[1- \sum_{i=1}^s \pm \Big(\nu_i \frac{\partial \tau}{\partial \rho_i}\Big) \bigg] \widetilde{\mathbf{Q}}-\widetilde{\mathbf{R}} .
\end{array}
\label{Eqcoef1}
\end{equation}
\end{theorem}

new

\begin{equation}
\begin{array}{l}
\left[\begin{array}{cc}
\dot{\mathbf{X}} + \mathbf{X} \mathbf{A} + \widehat{B}\mathbf{C}_2 + (\star)   & \star \\
\widehat{A}^{\text{T}} + \mathbf{A} + \mathbf{B}_2 \mathbf{D}_k \mathbf{C}_2 & -\dot{\mathbf{Y}} + \mathbf{A}\mathbf{Y} + \mathbf{B}_2 \widehat{C} + (\star) \\
(\mathbf{X} \mathbf{B}_1 + \widehat{B} \mathbf{D}_{21})^{\text{T}} & (\mathbf{B}_1 + \mathbf{B}_2 \mathbf{D}_k \mathbf{D}_{21})^{\text{T}} \\
\mathbf{C}_1 + \mathbf{D}_{12} \mathbf{D}_k \mathbf{C}_2 & \mathbf{C}_1 \mathbf{Y} + \mathbf{D}_{12} \widehat{C}
\end{array}\right.\\
\qquad\qquad\qquad\qquad\quad\quad\quad\qquad\left.\begin{array}{cc}
\star & \star \\
\star & \star \\
 -\gamma \mathbf{I}   & \star \\
\mathbf{D}_{11} + \mathbf{D}_{12} \mathbf{D}_k \mathbf{D}_{21} &  -\gamma \mathbf{I}
\end{array}\right]\prec\mathbf{0},
\end{array}
\end{equation}

\begin{equation}
\left[\begin{array}{cc}
\mathbf{X} & \mathbf{I} \\
\mathbf{I} & \mathbf{Y}
\end{array}\right] \succ	 \mathbf{0},
\end{equation}

\noindent Once the parameter dependent matrices $\mathbf{X}$, $\mathbf{Y}$, $\widehat{A}$, $\widehat{A}_d$, $\widehat{B}$, $\widehat{C}$, $\widehat{C}_d$, and $\widehat{D}_k$ satisfying the mentioned LMI condition (\ref{eq:LMI closed-loop}) have been obtained, the delayed output-feedback controller matrices can be computed through the following steps:

\noindent 1. Determine $\mathbf{M}$ and $\mathbf{N}$ from the factorization problem
\begin{equation}
\mathbf{I} - \mathbf{X}\mathbf{Y} = \mathbf{N} \mathbf{M}^{\text{T}},
\end{equation} 

\noindent where the obtained $\mathbf{M}$ and $\mathbf{N}$ matrices are always square and invertible in the case of computing full-order controllers.

\noindent 2. Compute the controller matrices by reversing the transformations defined by
\begin{equation}
\begin{array}{cl}
\widehat{A}&= \mathbf{X} \mathbf{A} \mathbf{Y} + \mathbf{X} \mathbf{B}_2 \mathbf{D}_k \mathbf{C}_2 \mathbf{Y} + \mathbf{N} \mathbf{B}_k \mathbf{C}_2 \mathbf{Y}\\
& + \mathbf{X} \mathbf{B}_2 \mathbf{C}_k \mathbf{M}^{\text{T}} + \mathbf{N} \mathbf{A}_k \mathbf{M}^{\text{T}},\\[0.2cm]
\widehat{A}_d&= \mathbf{X} \mathbf{A}_d \mathbf{Y} + \mathbf{X} \mathbf{B}_2 \mathbf{D}_k \mathbf{C}_{2d} \mathbf{Y} + \mathbf{N} \mathbf{B}_k \mathbf{C}_{2d} \mathbf{Y}\\
& + \mathbf{X} \mathbf{B}_2 \mathbf{C}_{dk} \mathbf{M}^{\text{T}} + \mathbf{N} \mathbf{A}_{dk} \mathbf{M}^{\text{T}},\\[0.2cm]
\widehat{B}&= \mathbf{X} \mathbf{B}_2 \mathbf{D}_k + \mathbf{N} \mathbf{B}_k,\\
\widehat{C}&= \mathbf{D}_k \mathbf{C}_2 \mathbf{Y} + \mathbf{C}_k \mathbf{M}^{\text{T}},\\
\widehat{C}_d&= \mathbf{D}_k \mathbf{C}_{2d} \mathbf{Y} + \mathbf{C}_{dk} \mathbf{M}^{\text{T}}.
\end{array}
\end{equation}

\noindent 3. Finally, the controller matrices are computed in the following order:
\begin{equation}
\begin{array}{cl}
\mathbf{C}_{dk}&= (\widehat{C}_d - \mathbf{D}_k \mathbf{C}_{2d} \mathbf{Y}) \mathbf{M} ^{-\text{T}},\\[0.20cm]
\mathbf{C}_k&= (\widehat{C} - \mathbf{D}_k \mathbf{C}_{2} \mathbf{Y}) \mathbf{M} ^{-\text{T}},\\[0.20cm]
\mathbf{B}_k&= \mathbf{N}^{-1} (\widehat{B} -  \mathbf{X} \mathbf{B}_2 \mathbf{D}_k),\\[0.20cm]
\mathbf{A}_{dk}&= -\mathbf{N}^{-1} (\mathbf{X} \mathbf{A}_d \mathbf{Y} + \mathbf{X} \mathbf{B}_2 \mathbf{D}_k \mathbf{C}_{2d} \mathbf{Y} + \mathbf{N} \mathbf{B}_k \mathbf{C}_{2d} \mathbf{Y}\\
&+  \mathbf{X} \mathbf{B}_2 \mathbf{C}_{dk} \mathbf{M} ^{\text{T}} - \widehat{A}_d) \mathbf{M} ^{-\text{T}},\\[0.20cm]
\mathbf{A}_{k}&= -\mathbf{N}^{-1} (\mathbf{X} A \mathbf{Y} + \mathbf{X} \mathbf{B}_2 \mathbf{D}_k \mathbf{C}_{2} \mathbf{Y} + \mathbf{N} \mathbf{B}_k \mathbf{C}_{2} \mathbf{Y}\\
&+ \mathbf{X} \mathbf{B}_2 \mathbf{C}_{k} \mathbf{M} ^{\text{T}} - \widehat{A}) \mathbf{M} ^{-\text{T}}.\\
\end{array}
\end{equation}
\section*{CLOSED-LOOP SIMULATION RESULTS OF MAP REGULATION USING LPV CONTROL}
\label{sec:num4}

The MAP dynamic regulation problem is formulated in an LPV framework utilizing the state equations in (\ref{LPV_MAP}) where the state-space matrices are as in (\ref{eq:matrices1}). Moreover, the vector of the target outputs to be controlled is $\mathbf{z} = [\phi x_e \:\: \xi u]^{\text{T}}$, \textit{i.e.} \begin{small}
$\mathbf{C}_{1}(\boldsymbol{\rho}(t))=\begin{bmatrix}
0 & 0 & \phi\\
0 & 0 & 0
\end{bmatrix},$
\end{small} $\mathbf{D}_{12}(\boldsymbol{\rho}(t))=[0, \:\: \xi]^{\text{T}}$. The matrices $\mathbf{C}_{1d}(\boldsymbol{\rho}(t))$ and $\mathbf{D}_{11}(\boldsymbol{\rho}(t))$ in (\ref{LPVsystem})  are zero matrices with compatible dimensions. The tracking error which is included in the state $x_e(t)$ and the control effort $u(t)$ are being penalized by the variables $\phi$ and $\xi$, respectively. The choice of the scalars $\phi$ and $\xi$ determines the relative weighting in the optimization scheme and depends on the desired performance objectives. The output-feedback controller is designed to minimize the induced $\mathscr{L}_2$ gain (or $\mathscr{H}_{\infty}$ norm) (\ref{eq:Performance Index}) of the closed-loop LPV system (\ref{eq:closed-loop system}). The design objective is to guarantee the closed-loop stability and minimize the worst case disturbance amplification over the entire range of model parameter variations. Theorem \ref{thm:thm3} is used to design an output-feedback controller which leads to an infinite-dimensional convex optimization problem with an infinite number of LMIs and decision variables. To overcome this challenge, we utilize the gridding approach introduced in \cite{apkarian1998advanced} to convert the infinite-dimensional problem to a finite-dimensional convex optimization problem. In this regard, we choose the functional dependence as $\mathbf{M}(\boldsymbol{\rho}(t))=\mathbf{M}_0 + \sum\limits_{i=1}^{s}\rho_i(t) \mathbf{M}_{i_1} + \frac{1}{2}\sum\limits_{i=1}^{s}\rho_i^2(t) \mathbf{M}_{i_2}$, where $\mathbf{M}(\boldsymbol{\rho}(t))$ represents any of the parameter-dependent matrices appearing in the LMI condition (\ref{eq:LMI closed-loop}). Finally, gridding the scheduling parameter space at appropriate intervals leads to a finite set of LMIs to be solved for the unknown matrices and $\gamma$. The MATLAB\textsuperscript{\tiny\textregistered} toolbox YALMIP is used to solve the introduced optimization problem \cite{lofberg2004yalmip}.

To evaluate the performance of the proposed LPV gain-scheduling output-feedback control design, collected animal experiment data is used to build a  patient's non-linear MAP response model based on (\ref{eq:MAP response TF}) where the instantaneous values of the model parameters $K$, $T$, and $\tau$ are generated as follows \cite{Luspay2015}.

\begin{itemize}

\item Sensitivity parameter, $K$: experiments have confirmed a regressive non-linear relationship between the vasoactive drug injection and the MAP response through which the patient's sensitivity decreases gradually on a constant vasoactive drug injection. This behavior is captured by the following non-linear relationship:
\begin{equation}
a_k \dot{K} + K  = k_0 exp\{ - k_1 i(t) \},
\label{eq:sensitivity}
\end{equation}
where $i(t)$ is the drug injection and $a_k$, $k_0$, and $k_1$ are uniformly distributed random coefficients based on Tab.~ \ref{tab:coef} \cite{Craig2004}. For example, a non-responsive patient to the injected vasoactive drug will be characterized by a low $k_0$ and a high $k_1$. 

\item Lag time, $T$: This parameter gradually increases with the injected drug volume and it can be modeled as:
\begin{equation}
T= sat_{\:[T_{\min}, T_{\max}]} \: \{b_{T} \int_{0}^{t} i(t) \:dt \},
\label{eq:lag}
\end{equation}
where $b_{T}$ is a uniformly distributed random variable which shows the inclination of the increase and varies based on Tab.~\ref{tab:coef}.
\item  Infusion delay, $\tau$: Based on the observations, the delay value has a peak shortly after the drug injection but it decays afterward. The following equation is used to describe the delay behavior:
\begin{equation}
\begin{cases}a_{\tau,2} \dddot{\tau} + a_{\tau,1} \ddot{\tau} + \dot{\tau} = b_{\tau,1} \dot{i}(t) + i(t), & t\geq t_{i_0}, \\\tau=0, & otherwise,\end{cases}
\label{eq:delay}
\end{equation}

\noindent where the saturation is imposed on the delay value, \textit{i.e.} $sat_{[\tau_{\min}, \tau_{\max}]} \: \tau$ and the uniformly distributed random variables $a_{\tau,2}$, $a_{\tau,1}$, and $b_{\tau,1}$ are listed in Tab.~\ref{tab:coef}.
\end{itemize}

\begin{table}[t]
\centering
\caption{Probabilistic distribution of the non-linear patient model coefficients with uniform distribution $\mathcal{U}$}
\label{tab:coef}
\begin{tabular}{cc}
\hline
Parameter    & Distribution                             \\ \hline
$a_k$        & $\mathcal{U}(500, 600)$                  \\
$k_0$        & $\mathcal{U}(0.1, 1)$                    \\
$k_1$        & $\mathcal{U}(0.002, 0.007)$              \\
$b_{T}$      & $\mathcal{U}(10^{-4}, 3 \times 10^{-4})$ \\
$a_{\tau,1}$ & $\mathcal{U}(5, 15)$                     \\
$a_{\tau,2}$ & $\mathcal{U}(5, 15)$                     \\
$b_{\tau,1}$ & $\mathcal{U}(80, 120)$                   \\ \hline
\end{tabular}
\end{table}

\begin{figure}[t] 
\hspace*{-.12in}
\centering \includegraphics[width=\columnwidth, height=2.4in]{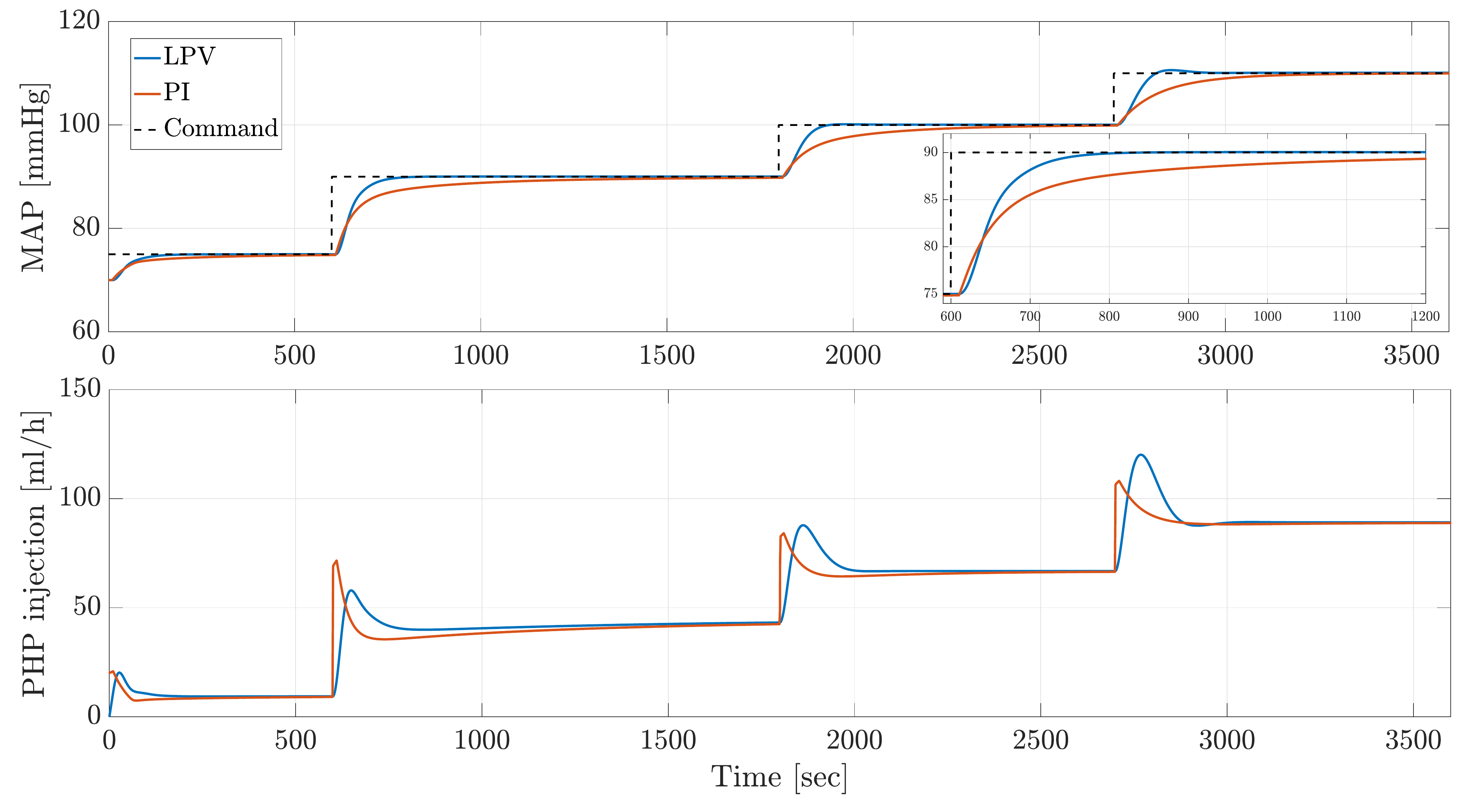}
\caption{Closed-loop MAP response and control effort of the proposed LPV controller and the fixed structure PI controller with no disturbance and no measurement noise} 
\label{fig:fig_tracking1}
\end{figure}

\begin{figure}[t] 
\hspace*{-.12in}
\centering \includegraphics[width=\columnwidth, height=2.1in]{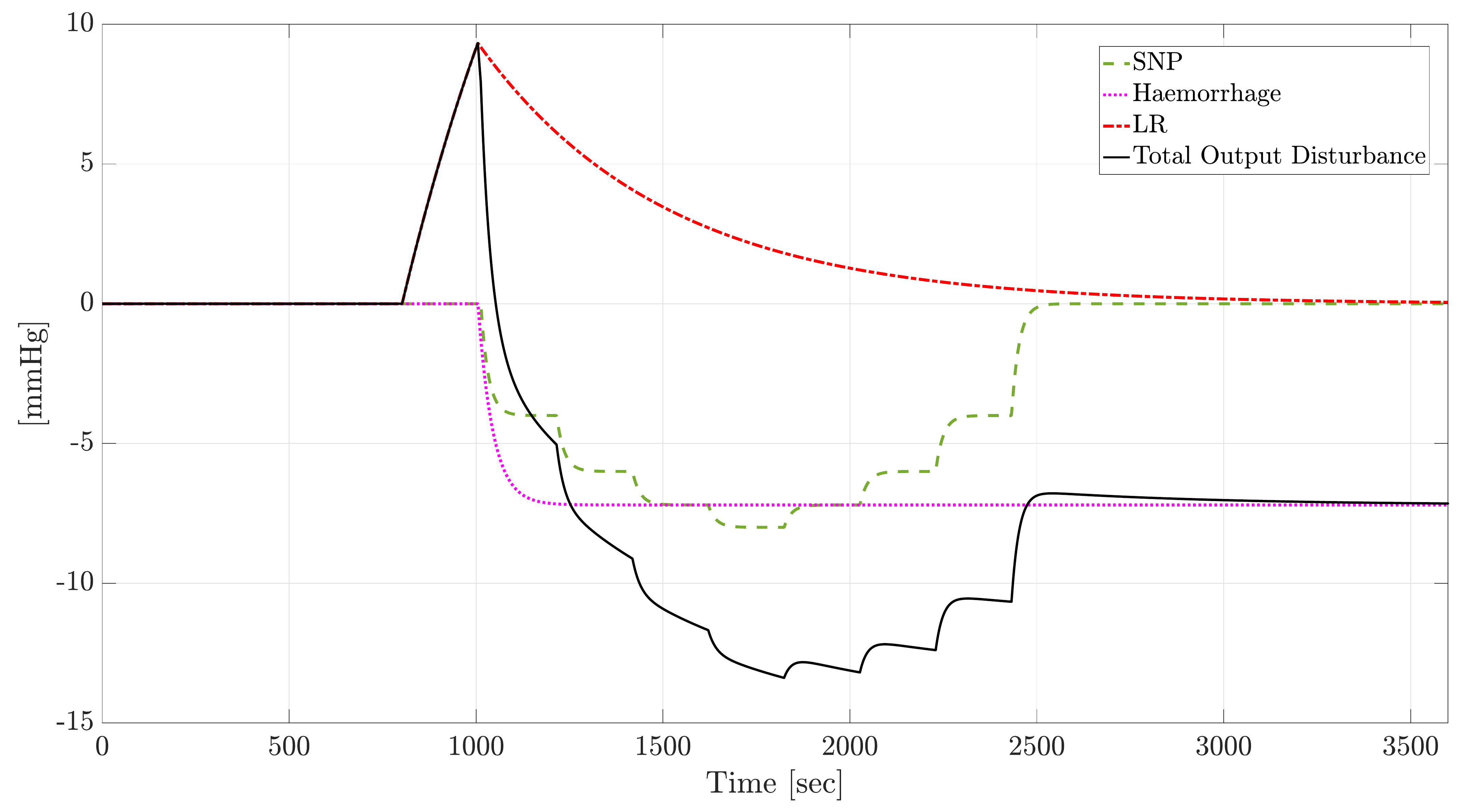}
\caption{Profile of output disturbances} 
\label{fig:Disturbance}
\end{figure}

\begin{figure}[t]
        \includegraphics[width=\columnwidth, height=2.4in]{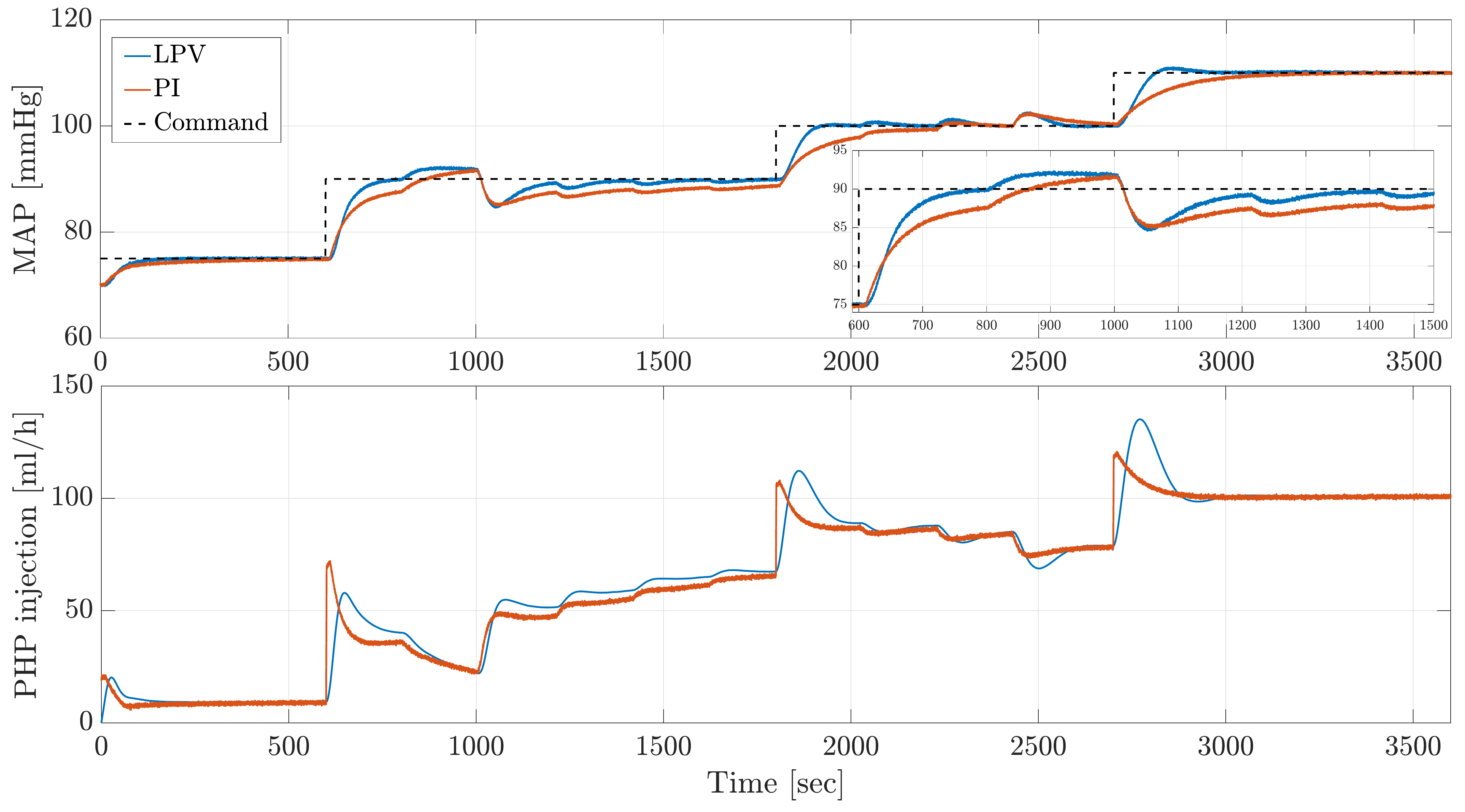}
    \caption{Closed-loop MAP response and control effort of proposed LPV controller against fixed structure PI controller subject to disturbance and measurement noise}
            \label{fig:trackingdistnoise}
\end{figure}

			For comparison, we evaluate the proposed controller performance against a fixed structure PI controller (see \cite{wassar2014automatic}). Given the following nominal values of the model parameters $\overline{K} = 0.55, \overline{T}=150$, and $\overline{\tau}=40$, the tuned PI controller transfer function is as follows:
		
        \begin{equation}
	     G_{c} (s) = 3 + \frac{0.017}{s},
		\end{equation}
	    
	   \noindent which is obtained based on the approach proposed in \cite{Zhong2006} to meet the prescribed gain and phase margin constraints. In the absence of disturbances and measurement noise, the tracking profile and the control effort are shown in Fig.~\ref{fig:fig_tracking1} where the objective is to regulate the MAP response to track the commanded MAP with minimum overshoot and the settling time and zero steady-state error. According to this figure, the overshoot of the closed-loop response remains within the admissible range and the delay-dependent parameter-varying controller provides a faster response with less settling time compared to the conventional PI controller. Next, we assume that the closed-loop system is experiencing both measurement noise and output disturbances. These disturbances could be the result of medical interventions and physiological variations due to hemorrhage or other medications like lactated ringers (LR). Figure~\ref{fig:Disturbance} is a typical profile of such disturbances. Considering measurement noise with the intensity of $10^{-3}$, the performance of the LPV and the PI  controllers can be seen in Fig.~\ref{fig:trackingdistnoise}. As expected, the proposed LPV controller outperforms the fixed structure PI controller with respect to the rise time and speed of the response due to its scheduling structure. To conclude, we can observe that the proposed LPV gain-scheduling control methodology with adaptation for wide range of patients demonstrates outstanding closed-loop performance in terms of MAP reference tracking and disturbance rejection under different scenarios and also is able to maintain the blood pressure within the allowable range of the reference value with maximum overshoot under $1\%$ and settling time under $200$ seconds. 
\section*{CONCLUSION AND FUTURE WORK}
\label{sec:num5}

Automated MAP feedback control for critical patient resuscitation is addressed in the present paper. In this regard, the governing dynamics of the MAP response to the vasopressor drug injection is expressed as an LPV time delayed system and a delay-dependent LPV output-feedback controller is designed to guarantee the asymptotic stability and the prescribed performance of the closed-loop system in terms of the induced $\mathcal{L}_2$-norm specification. It is shown that the utilized method and the choice of the parameter-dependent Lyapunov-Krasovskii functionals have resulted in a direct and also less conservative design approach, that can also handle the LPV systems with fast-varying parameters and time delay. Closed-loop simulations using a non-linear simulation patient model derived from experimental data demonstrated the controller desirable robustness to model parameters variations and time-delay while adjusting the set-point response under disturbances and measurement noise. Future research would focus on developing a new efficient online estimation algorithm to estimate the time-varying model parameters.

\vspace{-10mm}

\vspace{6mm}

\section*{ACKNOWLEDGMENT}\label{sec:Acknow}
Financial support from the National Science Foundation under grant CMMI1437532 is gratefully acknowledged. The collaboration of the Resuscitation Research Laboratory (Dr. G. Kramer) at the University of Texas Medical Branch (UTMB), Galveston, Texas, in providing animal experiment data is gratefully acknowledged.

\vspace{-1mm}

\bibliographystyle{asmems4}
\bibliography{asme2e.bib}


%
\end{document}